\documentclass{article}
\usepackage{spconf,amsmath,graphicx}

\usepackage{enumitem}
\setlist{nosep, leftmargin=14pt}
\usepackage{bbm}
\usepackage{mwe} % to get dummy images
\usepackage[hidelinks]{hyperref}
\usepackage{booktabs}
\usepackage{multirow}
\usepackage{colortbl}
\usepackage[T1]{fontenc}
\usepackage{array}
\usepackage{tabularray}
\title{Automatic lobe segmentation using attentive cross entropy and end-to-end fissure generation}
\name{Qi Su$^{1,2}$ \qquad Na Wang$^{3}$ \qquad Jiawen Xie$^{1}$ \qquad Yinan Chen$^{3,4}$ \qquad Xiaofan Zhang$^{1,2}$}

\address{ $^{1}$ Qing Yuan Research Institute, Shanghai Jiao Tong University \\
$^{2}$ Shanghai AI Laboratory \qquad 
$^{3}$ SenseTime Research \\
$^{4}$ West China Biomedical Big Data Center, Sichuan University West China Hospital
}
\begin{document}
%\ninept
%
\maketitle
\begin{abstract}
The automatic lung lobe segmentation algorithm is of great significance for the diagnosis and treatment of lung diseases, however, which has great challenges due to the incompleteness of pulmonary fissures in lung CT images and the large variability of pathological features. Therefore, we propose a new automatic lung lobe segmentation framework, in which we urge the model to pay attention to the area around the pulmonary fissure during the training process, which is realized by a task-specific loss function. In addition, we introduce an end-to-end pulmonary fissure generation method in the auxiliary pulmonary fissure segmentation task, without any additional network branch.  Finally, we propose a registration-based loss function to alleviate the convergence difficulty of the Dice loss supervised pulmonary fissure segmentation task. We achieve 97.83\% and 94.75\% dice scores on our private dataset STLB and public LUNA16 dataset respectively.
\end{abstract}
\section{Introduction}
\label{sec:intro}

The human lung is composed of five lobes separated by pulmonary fissures, in which the left lung is divided into left upper and lower lobes by a large left oblique fissure (LOF), while the right lung is divided into right upper, middle, and lower lobes by a large right oblique fissure (ROF) and a small right horizontal fissure (RHF). Each pulmonary lobe is independent in physiological function, which makes it becomes the unit where many diseases occur and spread. Therefore, it is instrumental for medical diagnosis to separate each pulmonary lobe for analysis.

In lung computed tomography (CT) images, the pulmonary fissure is a thin sheet structure that has a large Hounsfiled Unit (HU) value relative to the surrounding lung tissue. The lobe segmentation task will be solved easily if the location of the pulmonary fissure can be found accurately. However, the pulmonary fissure is partially incomplete or even completely invisible in many cases \cite{raasch1982radiographic}. On this condition, it will be an intractable task to segment each lobe with the human naked eyes, which will consume much manpower and time. Therefore, the automatic lung lobe segmentation method is particularly important.

In previous works, Harrison \textit{et al.} \cite{harrison2017progressive} enhance the vanilla CNN model with the progressive multi-path scheme which integrates the outputs from different depths of the network. Ferreira \textit{et al.} \cite{ferreira2018end} use a 3D fully convolutional neural network called Fully Regularized V-Net, which involves various regularization methods, to perform lobe prediction. PDV-Net \cite{imran2018automatic} designs an end-to-end progressive dense V-Network that combines dense V-Net \cite{gibson2018automatic} and the work of \cite{harrison2017progressive}. FissureNet \cite{gerard2018fissurenet} uses two cascaded CNN to predict pulmonary fissure from coarse to fine, instead of directly predicting the lobe. RTSU-Net \cite{xie2020relational} introduces a non-local module to capture the structural relationship between lung lobes. In this paper, we propose several targeted methods to optimize the lobe segmentation task. First, we design a modified Cross Entropy loss with dynamic attention throughout the training process, we call it Attentive Cross Entropy loss. Second, we introduce an end-to-end pulmonary fissure generation method in the auxiliary fissure segmentation task, without adding any additional network branch. Finally, we use a registration-based loss as the optimization target of the auxiliary task to alleviate the convergence difficulty of Dice loss in this task. On our private STLB and the public LUNA16 dataset, we achieve 97.83\% and 94.75\% dice scores respectively, which have a great improvement compared with the baseline.

\section{Methodology}
\label{sec:format}

\begin{figure}[htb]
\begin{minipage}[b]{\linewidth}
  \centering
  \centerline{\includegraphics[width=8.1cm]{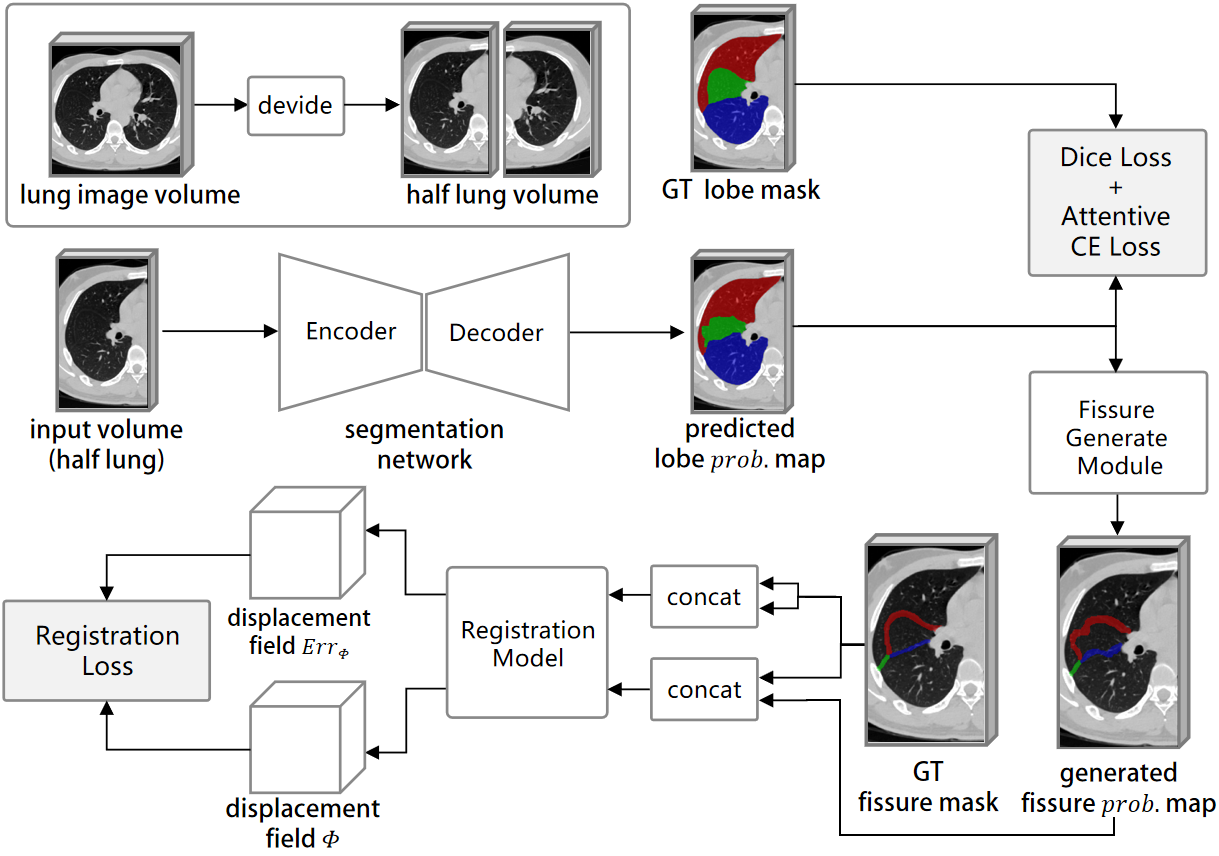}}
\end{minipage}
\caption{The pipeline of our proposed method. The input volume first goes through a segmentation network and produces the lobe probability map. The Attentive Cross Entropy loss and Dice loss are then calculated on the predicted lobe probability map and its ground truth mask. Thereafter, the fissure probability map is generated by the fissure generation module (FGM) with the lobe probability map as input. Finally, the generated fissure probability map and its ground truth mask are together used to be the inputs of the registration network to obtain two displacement fields $\phi$ and $Err_{\phi}$, so as to obtain the registration loss.}
\label{fig:frame}
\end{figure}

The pipeline of the proposed method is shown in \hyperref[fig:frame]{figure 1}. We will discuss this series of strategies in detail.

\subsection{Attentive Cross Entropy loss}
\label{ssec:ace}

The precision of lung lobe segmentation largely depends on the accurate segmentation of the area around the pulmonary fissure. However, the boundary between each lung lobe is hard to capture, which is the major difficulty of the lobe segmentation task. Therefore, it's natural to think of making the model pay attention to the voxels around the pulmonary fissure, that is, the hard voxels. Inspired by anchor loss \cite{ryou2019anchor} which assigns higher loss weight to hard cases, we change the loss weight of each voxel to achieve a similar purpose. Specifically, for a lung CT image $X$, we give each voxel $x_i$ ($x_i\in X$) a weight $w_i$ defined as follows:
\begin{align}
    w_i&=1-y_{i,g\left(x_i\right)}\cdot\alpha\\
       &=\left(1-y_{i,g\left(x_i\right)}\right)\cdot\alpha + 1\cdot\left(1-\alpha\right)
\end{align}
Suppose that the number of lung lobe foreground classes is $C_{\rm lobe}$, then $g(x_i)\in \{1,\dots,C_{\rm lobe}\}$ is the ground truth class of voxel $x_i$, $y_{i,g(x_i)}$ is the produced softmax probability that voxel $x_i$ belongs to class $g(x_i)$. $\alpha\in[0,1]$ is a hyper-parameter that controls the degree of attention to the misclassified voxels. It is worth noting that $\alpha$ should be a dynamic value gradually increasing from zero in the whole training process because at the initial stage of training the model has not yet converged, thus the so-called misclassified voxels may be distributed randomly in the volume which will make the model collapse if $\alpha$ is set to a large value at this time. Based on the above analysis, we define our Attentive Cross Entropy loss $L_{ace}$ with dynamic attention as follows:
\begin{equation}
L_{ace}=-\frac{1}{|\mathcal{D}|}\sum_{X\in \mathcal{D}}\sum_{x_i\in X}w_i{\rm{log}}\ y_{i,g(x_i)}
\end{equation}
in which $\mathcal{D}$ is the whole training dataset and $|\mathcal{D}|$ is the training dataset size.

\subsection{Auxiliary pulmonary fissure segmentation}
\label{ssec:subhead}

\subsubsection{Generation of fissure ground truth mask}
\label{sssec:subsubhead}

We generate the pulmonary fissure mask from the lobe mask as the ground truth of the auxiliary fissure segmentation task. Specifically, we perform morphological dilation on the binary masks of the two adjacent lobes of a specific fissure respectively and treat the intersection of the two dilated masks as the corresponding fissure mask. It is worth noting that we can get three fissure foreground classes on the right lung with this method, rather than two in anatomy.

\subsubsection{Gradient traceable fissure generation module}
\label{sssec:subsubhead}

Some works \cite{ferreira2018end,xie2020relational} add an additional network branch to predict the pulmonary fissure. However, the supervisory signal from the fissure can only be transmitted to a part of the network in this way. In addition, we find in our experiments that the extra branch often obtains poor fissure segmentation, even though the lobes can be well segmented. We suppose that it is owing to the hardship for the backbone before the two branches to extract features that are effective enough for both lobe and fissure segmentation tasks, which makes the model performing better on the simpler lobe segmentation task, while poorly on the more difficult fissure segmentation task. In this regard, we adopt an end-to-end approach to obtain lung fissure prediction without involving additional network branches.

\begin{figure}[tb]
\begin{minipage}[b]{1.0\linewidth}
\centering
\centerline{\includegraphics[width=7cm]{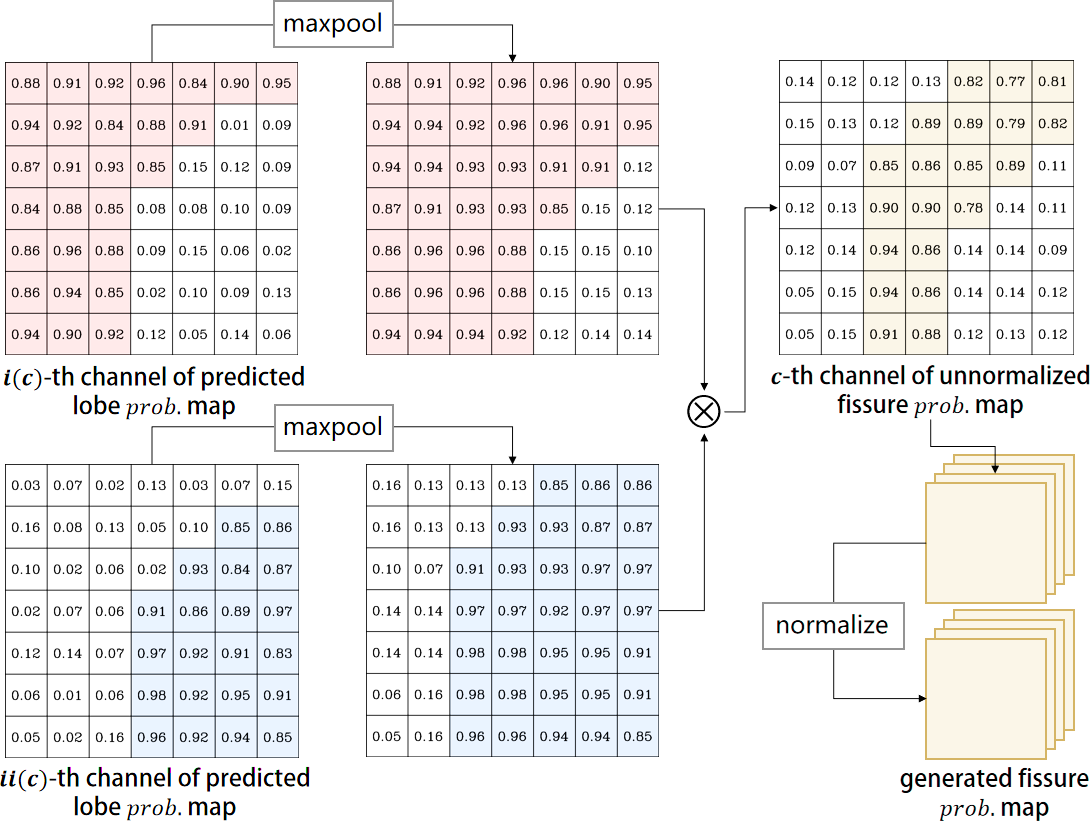}}
\end{minipage}
\caption{The workflow of the fissure generation module (FGM). In order to obtain the probability map of the $c$-th fissure foreground class, we perform maxpool on the predicted probability maps of its two adjacent lobe classes $i(c)$ and $ii(c)$, and then multiply the two element-wise. After performing on each fissure foreground class, we normalize the result so that the probabilities of all classes for one voxel sum to 1.}
\label{fig:fgm}
\end{figure}

We add a fissure generation module (FGM) which is inspired by the way we synthesize the fissure ground truth mask, after the last softmax layer. The fissure generation module uses a gradient traceable method to obtain the prediction of the pulmonary fissure, its core processing flow is shown in \hyperref[fig:fgm]{figure 2}. Specifically, suppose that the predicted lung lobe probability map is $Y$ for the input image $X$, and $Y_c$ is the $c$-th channel of $Y$ which represents the probability that $X$ belongs to lobe class $c$. We use the following formula to obtain the predicted fissure probability map $Z$:
\begin{align}
    \hat{Z}_c&={\rm{mp}}\left(Y_{\expandafter{\romannumeral1}(c)}\right)\cdot {\rm{mp}}\left(Y_{\expandafter{\romannumeral2}(c)}\right),c=1,2,\dots,C_{\rm fiss}\\
    \hat{Z}_0&=\prod_{c=1}^{C_{\rm fiss}}(1-\hat{Z}_c) \label{4}\\
    Z_c&=\frac{\hat{Z}_c}{\sum_{i=0}^{C_{\rm fiss}} \hat{Z}_i},c=0,1,\dots,C_{\rm fiss} \label{5}
\end{align}
in which $\expandafter{\romannumeral1}(c)$, $\expandafter{\romannumeral2}(c)\in \{1,\dots,C_{\rm lobe}\}$ represent the two lobar classes adjacent to fissure class $c$, and $C_{\rm fiss}$ is the total number of pulmonary fissure classes. The function ${\rm{mp}}(\cdot)$ indicates max pooling operation, which is used to simulate the morphological dilation, meanwhile, the voxel-wise multiplication simulates intersection operations, with gradient traceable. After that, \eqref{4} simply multiplies the antiphase of all foreground probability maps as the counterpart of the background. Finally, \eqref{5} normalizes the probability map in the channel direction. All operations of the fissure generation module are derivable so that the supervisory signal of the lung fissure can be transmitted back to the whole network.

\subsection{Registration loss for pulmonary fissure}
\label{ssec:regis}

The pulmonary fissure is quite a thin structure on which the Dice loss \cite{milletari2016v} is prone to oscillate and difficult to optimize. In addition, Dice loss cannot well constrain the shape and structure of the pulmonary fissure. Therefore, we propose the registration loss as an alternate.

Specifically, we train a registration model $\phi(\cdot,\cdot)$ for the pulmonary fissure, which outputs the displacement field of the two input fissure masks. A robust registration model will capture the geometric information of the two inputs, and feed them back into the output displacement field. When the generated fissure is similar to the corresponding ground truth mask, the norm of the displacement field should be small, and vice versa. Therefore, we define the registration loss as:
\begin{align}
    &L_{reg}=\frac{1}{|\mathcal{D}|}\sum_{X\in \mathcal{D}}||\phi\left(Z,G_f(X)\right)-Err_{\phi}(X)||_1\\
    &Err_{\phi}(X)=\phi\left(G_f(X),G_f(X)\right)
\end{align}
in which $G_f(X)$ is the fissure ground truth mask of sample $X$. The involvement of $Err_{\phi}(X)$ is to offset the error of the registration model itself.

\subsection{Learning objective}
\label{ssec:subhead}
For the lung lobe, we also use ordinary Dice loss \cite{milletari2016v} $L_{dc}$ in addition to $L_{ace}$ mentioned in \hyperref[ssec:ace]{section 2.1}, so the final learning object is defined as:
\begin{equation}
    L=\lambda_1 L_{ace}+\lambda_2 L_{dc}+\lambda_3 L_{reg}
\end{equation}
$\lambda_{1}$, $\lambda_{2}$ and $\lambda_{3}$ are hyper-parameters which control the weight of each loss. In our implementation, we fix $\lambda_1$ and $\lambda_2$ to 1 and increase $\lambda_3$ linearly from 0 to 1 with the training progressing.

\section{Experiment}
\label{sec:pagestyle}

\subsection{Datasets and preprocessing}
\label{ssec:subhead}

We validate our method on two lung CT datasets. First is our private lung dataset STLB, we randomly select a subset of 207 for our experiment, in which 123 cases are for training, 42 cases are for validation, and the rest is for testing. Besides, we also use a subset of the public LUNA16 \cite{setio2017validation} dataset with a size of 51, the annotations are provided by Tang \textit{et al.} \cite{tang2019automatic}, we use 41 samples for training and 10 for testing. In terms of pathology, our STLB dataset contains a small part of samples with pulmonary nodules, tuberculosis, and pneumonia which will affect the visibility of the fissures, and the LUNA16 dataset contains a large number of samples with pulmonary nodules. In practice, we find that our method is insensitive to these pathological features as long as the fissures are relatively clear. In the preprocessing stage, we uniformly resample the image volume to $1mm\times 1mm\times 1mm$ spacing, and normalize the voxel value within HU window $[-1000,400]$ to $[0,1]$.

\begin{center}
\begin{figure*}[htb]
\centering
\includegraphics[width=0.80\linewidth]{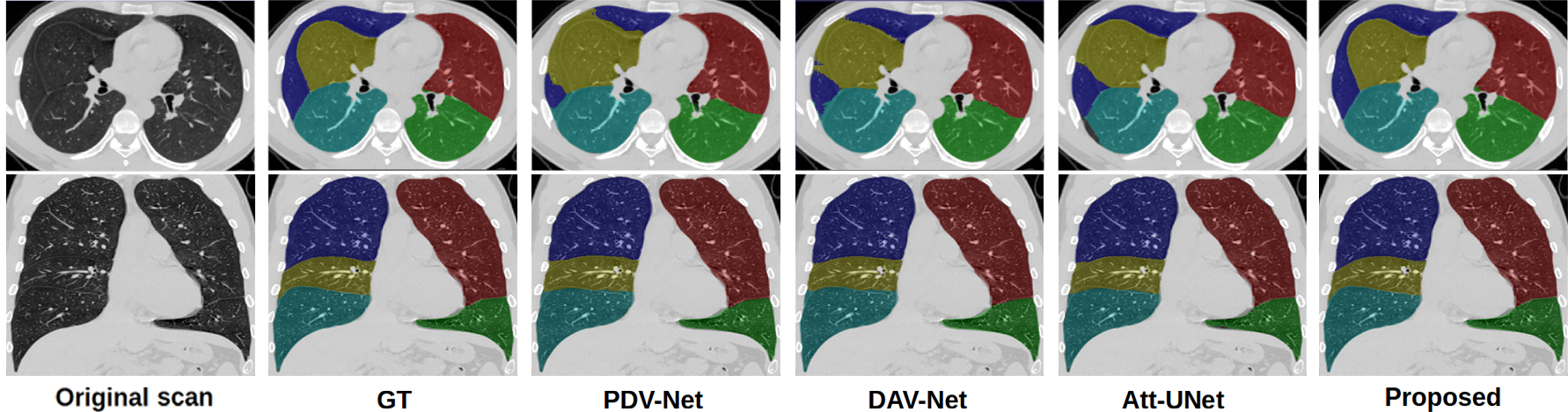}
\caption{Visualization results of each method. The two lines are the slices from the cross section and coronal plane respectively. We can see that our method can achieve more accurate segmentation results near the pulmonary fissure than other methods.}
\label{fig:segment}
\end{figure*}
\end{center}

\subsection{Implementation}
\label{ssec:subhead}

We implement our method using Pytorch 1.10.0 with one NVIDIA GeForce RTX 3090 GPU. We use vanilla 3D U-Net \cite{cciccek20163d} as the backbone of our segmentation model. The network is trained using Adam \cite{kingma2014adam} optimizer with a constant learning rate of 0.001 and weight decay of $10^{-4}$. All the training experiments have experienced $10^4$ epochs, no extra data augmentation methods are used except for random noise. During training, we directly use the complete half lung volume as the model input, meanwhile setting the batch size to 1. Using a complete lung for training is also feasible, we use half a lung volume for saving GPU memory. Finally, we keep the optimal model on the validation set, and calculate the mean dice score (DSC) and Hausdorff distance (95\%) (HD95) on the test set as the evaluation metrics. For the registration model mentioned in \hyperref[ssec:regis]{section 2.3}, we implement it with VoxelMorph \cite{balakrishnan2019voxelmorph}, an effective registration method for medical images.

\subsection{Results and analysis}
\label{ssec:results}

We compare our results with some previous lobe segmentation methods, the results are in the \hyperref[tab1]{table 1}. On the STLB dataset, our results have significantly improved compared with other methods both on DSC and HD95 metrics. Especially, we have achieved a 2.40\% higher dice score than the second place \cite{oktay2018attention} on the right middle lobe, whose accurate segmentation is supposed to be the toughest of the five lung lobes. On the smaller LUNA16 dataset, we find that all the methods achieve worse results than those obtained on the larger STLB dataset, nevertheless, our proposed method is still slightly prominent in all the comparative experiments. We also qualitatively show the segmentation results of each method in \hyperref[fig:segment]{figure 3}. The boundary of the lobe, especially the right middle lobe, is smoother and matches the real fissure location better in our result.

\begin{center}
\begin{table}[htb]
\centering{
    \scalebox{0.53}{
        \begin{tabular}{ccccccc} 
\toprule
                    \#STLB                         & Left-Upper                       & Left-Lower                      & Right-Upper                      & Right-Mid                        & Right-Low                       & Mean                            \\ 
\midrule
\multirow{2}{*}{PDVNet \cite{imran2018automatic}}                      & 97.26$\pm$1.73                     & 96.94$\pm$1.75                     & 95.71$\pm$1.86                     & 92.20$\pm$3.54                     & 97.17$\pm$0.91                     & 95.85$\pm$2.87                      \\
                                              & 3.98$\pm$5.69                      & 4.23$\pm$4.89                      & 7.46$\pm$7.22                      & 7.95$\pm$5.59                      & 3.37$\pm$2.65                      & 5.40$\pm$5.74                       \\
\multirow{2}{*}{DAVNet \cite{zheng2021dual}}                      &       97.00$\pm$2.25               &         96.85$\pm$2.43             &        95.95$\pm$2.57              &         92.90$\pm$3.60             &       97.60$\pm$1.11               &           96.06$\pm$3.02            \\
                                              &           5.07$\pm$6.22            &       6.06$\pm$12.01                &       7.54$\pm$6.86                &        7.72$\pm$6.06               &         3.21$\pm$3.33              &         5.92$\pm$7.64               \\
\multirow{2}{*}{Att-UNet \cite{oktay2018attention}}                      &          97.32$\pm$1.95            &          96.59$\pm$3.05            &         96.44$\pm$1.68             &         93.84$\pm$2.83             &          97.72$\pm$0.88            &      96.38$\pm$2.60                 \\
                                              &         4.93$\pm$8.07              &          5.22$\pm$6.36             &       5.78$\pm$3.83                &          6.44$\pm$4.67             &         3.06$\pm$3.34              &          5.09$\pm$5.65              \\
\multicolumn{1}{l}{\multirow{2}{*}{Proposed}} & \multicolumn{1}{l}{\textbf{98.36$\pm$1.47}} & \multicolumn{1}{l}{\textbf{98.24$\pm$1.61}} & \multicolumn{1}{l}{\textbf{97.77$\pm$1.87}} & \multicolumn{1}{l}{\textbf{96.23$\pm$2.87}} & \multicolumn{1}{l}{\textbf{98.57$\pm$0.75}} & \multicolumn{1}{l}{\textbf{97.83$\pm$2.06}}  \\
 & \multicolumn{1}{l}{\textbf{2.14$\pm$3.46}}  & \multicolumn{1}{l}{\textbf{2.31$\pm$3.92}}  & \multicolumn{1}{l}{\textbf{3.37$\pm$3.77}}  & \multicolumn{1}{l}{\textbf{4.78$\pm$5.38}}  & \multicolumn{1}{l}{\textbf{1.56$\pm$1.58}}  & \multicolumn{1}{l}{\textbf{2.83$\pm$3.99}}   \\
\bottomrule
\\
\toprule
                \#LUNA16                  & Left-Upper                       & Left-Lower                      & Right-Upper                      & Right-Mid                        & Right-Low                       & Mean                            \\ 
\midrule
\multirow{2}{*}{PDV-Net \cite{imran2018automatic}}                      & 95.02$\pm$3.11                     & 95.06$\pm$2.66                     & 93.20$\pm$2.86                     & 87.47$\pm$3.95                     & 95.66$\pm$1.63                     & 93.28$\pm$4.21                      \\
                                              & 7.93$\pm$7.56                      & 8.63$\pm$7.35                      & 10.68$\pm$4.13                      & 15.47$\pm$6.32                      & 8.36$\pm$7.35                      & 10.21$\pm$7.23                       \\
\multirow{2}{*}{DAV-Net \cite{zheng2021dual}}                      & 96.87$\pm$1.65                     & 93.34$\pm$7.22                     & 93.46$\pm$3.54                     & 85.05$\pm$5.06                     & 95.74$\pm$1.76                     & 92.89$\pm$6.03                      \\
                                              & 5.80$\pm$5.32                      & 9.31$\pm$10.34                      & 17.24$\pm$8.30                      & 15.52$\pm$3.89                      & 7.83$\pm$3.94                      & 11.14$\pm$8.17                       \\
\multirow{2}{*}{Att-UNet \cite{oktay2018attention}}                      & 96.00$\pm$3.74                     & 94.38$\pm$5.11                     & \textbf{93.98$\pm$2.40}                     & 87.72$\pm$3.38                     & 95.89$\pm$1.64                     & 93.59$\pm$4.61                      \\
                                              & 6.36$\pm$7.64                      & 8.72$\pm$9.18                      & \textbf{10.12$\pm$3.56}                      & 16.24$\pm$5.44                      & 6.42$\pm$2.71                      & 9.57$\pm$7.18                       \\
\multicolumn{1}{l}{\multirow{2}{*}{Proposed}} & \multicolumn{1}{l}{\textbf{97.69$\pm$1.59}} & \multicolumn{1}{l}{\textbf{97.38$\pm$1.54}} & \multicolumn{1}{l}{93.90$\pm$4.34} & \multicolumn{1}{l}{\textbf{88.10$\pm$5.94}} & \multicolumn{1}{l}{\textbf{96.67$\pm$1.77}} & \multicolumn{1}{l}{\textbf{94.75$\pm$5.03}}  \\
\multicolumn{1}{l}{}                          & \multicolumn{1}{l}{\textbf{4.18$\pm$5.67}}  & \multicolumn{1}{l}{\textbf{3.86$\pm$4.53}}  & \multicolumn{1}{l}{10.47$\pm$6.38}  & \multicolumn{1}{l}{\textbf{14.12$\pm$5.24}}  & \multicolumn{1}{l}{\textbf{7.18$\pm$8.44}}  & \multicolumn{1}{l}{\textbf{7.96$\pm$7.32}}   \\
\bottomrule
\end{tabular}
    }
}
\caption{The results of our proposed method and some mainstream methods on both STLB and LUNA16 datasets. The upper result on each line is DSC(\%), and the lower is HD95.}
\label{tab1}
\end{table}
\end{center}

\subsection{Ablation Studies}
\label{ssec:subhead}

In order to verify the effectiveness of each module we proposed, we conduct a series of ablation studies using the STLB dataset. First, we train a vanilla 3D U-Net \cite{cciccek20163d} model without any proposed method as the baseline, then we merely add our Attentive Cross Entropy loss to validate its effect, and finally, we respectively use the Dice loss and Registration loss as the training objective for the generated fissure in the auxiliary task. The results of all ablation experiments are in the \hyperref[tab2]{table 2}. From the table, the obtained results using our proposed Attentive Cross Entropy loss for the lobe segmentation task are significantly improved in dice score compared with baseline. On this premise, when using Dice loss as the supervisory signal in the fissure segmentation auxiliary task to supervise the generated pulmonary fissure, the dice score decreases instead. In fact, we find that the Dice loss in this auxiliary task is constantly fluctuating and nonconvergent during training. After using our registration loss instead, the DSC and ASSD metrics are both improved, especially on the right middle lobe and the small right horizontal fissure respectively.

\begin{center}
\begin{table}[h]
\centering{
    \scalebox{0.53}{
\begin{tblr}{
  cells = {c},
  cell{6}{6} = {c=2}{},
  cell{7}{6} = {c=2}{},
  cell{8}{6} = {c=2}{},
  cell{9}{6} = {c=2}{},
  hline{1,6,10} = {-}{0.08em},
  hline{2,7} = {-}{0.05em},
}
DSC(\%)  & Left-Upper & Left-Lower & Right-Upper & RIght-Mid  & Right-Lower & Mean       \\
Baseline  & 97.72$\pm$1.76 & 97.55$\pm$1.73 & 97.08$\pm$1.78  & 94.86$\pm$3.24 & 98.14$\pm$0.84  & 97.07$\pm$2.33 \\
w/ACE      & 98.35$\pm$1.72 & 98.24$\pm$1.63 & 97.69$\pm$1.78  & 95.99$\pm$3.06 & 98.51$\pm$0.77  & 97.76$\pm$2.14 \\
w/ACE\&Dice & 98.17$\pm$2.05 & 98.08$\pm$1.88 & 97.61$\pm$2.00  & 95.97$\pm$3.50 & 98.47$\pm$0.81  & 97.66$\pm$2.39 \\
w/ACE\&Reg  & \textbf{98.36$\pm$1.67} & \textbf{98.24$\pm$1.61} & \textbf{97.77$\pm$1.87}  & \textbf{96.23$\pm$2.87} & \textbf{98.57$\pm$0.75}  & \textbf{97.83$\pm$2.06} \\
ASSD      & LOF        & RHF        & ROF-Upper   & ROF-Lower  & Mean        &            \\
w/ACE      & 1.88$\pm$2.00  & 2.66$\pm$2.35  & 1.90$\pm$2.43   & 1.61$\pm$0.98  & 2.01$\pm$2.06   &            \\
w/ACE\&Dice & 1.97$\pm$2.62  & 2.50$\pm$2.43  & 1.77$\pm$2.71   & 1.81$\pm$0.92  & 2.01$\pm$2.30   &            \\
w/ACE\&Reg  & \textbf{1.78$\pm$2.18}  & \textbf{2.30$\pm$2.25}  & \textbf{1.73$\pm$2.28}   & \textbf{1.44$\pm$0.94}  & \textbf{1.81$\pm$2.06}   &            
\end{tblr}
    }
}
\caption{Results of ablation studies on STLB dataset. ACE denotes our Attentive Cross Entropy loss, and Dice and Reg denote conducting the auxiliary fissure segmentation task with Dice loss or Registration loss as the learning objective. We evaluate each method with both the Dice scores of the five lobes and the average symmetric surface distances (ASSD) along the four fissures.}
\label{tab2}
\end{table}
\end{center}

\section{CONCLUSIONS}
\label{sec:majhead}

In this paper, we propose a novel lung lobe segmentation pipeline, which includes a task-specific Attentive Cross Entropy loss that allows the model to constantly pay attention to the area around the pulmonary fissure during training, and an end-to-end method to generate fissure prediction as the auxiliary task, with a designed registration-based loss for it. We evaluate our method on two lung CT image datasets and achieve satisfactory results on both datasets, which indicates the great robustness and clinical application value of our method.

\bibliographystyle{IEEEbib}
\bibliography{strings,refs}

\end{document}